\documentclass[3p,times,twocolumn]{elsarticle}
 \biboptions{comma,sort&compress}
 
\usepackage{graphicx}
\usepackage{slashed}
\usepackage{amsmath}
\usepackage{amssymb}
\usepackage{here}

\usepackage{float}
\graphicspath{{plots/}}

\usepackage{physics}
\usepackage{tikz}
\usepackage{tikz-feynman}
\usepackage[caption = false]{subfig}
\usepackage{enumerate}
\usepackage{enumitem}
\usepackage{siunitx}
\usepackage{ecrc}

\volume{00}

\firstpage{1}

\journalname{Nuclear and Particle Physics Proceedings}

\runauth{}

\jid{nppp}

\jnltitlelogo{Nuclear and Particle Physics Proceedings}

\usepackage{amssymb}

\usepackage[figuresright]{rotating}

\begin{document}

\begin{frontmatter}

\title{$\pi$-$\pi$ scattering lengths: electric corrections in the linear sigma model}
 \cortext[cor0]{Talk given at 26th International Conference in Quantum Chromodynamics (QCD 24), Montpellier - FR}

 \author[label2,label3]{M. Loewe \fnref{fn1}}
  \fntext[fn1]{Speaker, Corresponding author.}
 \address[label2]{Facultad de Ingenier\'ia, Arquitectura y Diseño, Universidad San Sebasti\'an, Santiago, Chile.}
\address[label3]{Centre for Theoretical and Mathematical Physics, and Department of Physics, University of Cape Town, Rondebosch 7700, South Africa.}

\ead{mloewelo@yahoo.com}

 \author[label4]{R. Cádiz}
  

\address[label4]{Department of Physics and Astronomy, Stony Brook University, Stony Brook, NY 11794, USA.}
 \author[label2,label5]{R. Zamora}
 \address[label5]{Instituto de Ciencias B\'asicas, Universidad Diego Portales, Casilla 298-V, Santiago, Chile.}

\pagestyle{myheadings}
\markright{ }
\begin{abstract}
 In the frame of the linear sigma model and working in the weak field approximation, we discuss the role played by an external electric field on the behavior of $\pi$-$\pi$ scattering lengths. For this purpose,  we have considered all relevant one-loop diagrams in the $s$, $t$, and $u$ channels where we have Schwinger propagators for charged pions. An important novelty of our analysis, compared with previous existing discussions in the literature, is the explicit calculation of box diagrams which were previously not considered in discussions regarding magnetic corrections to  $\pi$-$\pi$ scattering lengths. Our analysis shows that electric field corrections have an opposite effect with respect to the previously calculated magnetic corrections.
\end{abstract}
\begin{keyword}  
Linear Sigma model; External electromagnetic fields; Pion-Pion scattering lengths

\end{keyword}

\end{frontmatter}
\section{Introduction}
The behavior of matter under extreme conditions, due to external agents, have attracted the attention of the community during the last years. Perhaps the most important scenario where such effects appear corresponds to relativistic heavy ion collisions, like Au-Au, where extreme high temperatures are produced inducing the occurrence of  different phase transitions as deconfinement or chiral symmetry restoration. The inclusion of density effects can also be taken into account, emerging then  a rich phase diagram in the temperature-density plane. The  latter scenario corresponds to extreme  compact objects like neutron stars. During the initial stages of of the collision between heavy nuclei, like Au-Au,  an extremely high magnetic field is produced . The case of relativistic collisions between a heavy and a light nuclei, for example Au-Cu, is also very interesting due to the appearance of an electric field, due to the imbalance in the number of protons associated to each nucleus. Both types  fields, electric and the magnetic, are produced in a relative perpendicular configuration, being the magnetic field produced in a direction essentially perpendicular to the collision plane, whereas the electric field points along the collision plane.  In the existing literature, there are several works related to the study of different physical parameters in the presence of a magnetic and/or electric field \cite{Johnson:2008vna,Frasca:2011zn,Bandyopadhyay:2020zte,G1,Zayakin:2008cy,Filev:2009xp,Ballon-Bayona:2020xtf,G2,G3,G4,G5,fernandez,Hernandez:2024nev}.

It is important to elucidate  the relative effects of the external agents on measurable  physical quantities. For example, we know that temperature and magnetic field conspire again each other in several scenarios. Here we consider the effect of an external weak electric field on $\pi $-$\pi $ scattering lengths, being these pions produced during the collision.

\smallskip 
For this purpose we work in the frame of the linear sigma model, computing all relevant diagrams. Fermion contributions are neglected in the analysis since they are much more massive than the pions and the scalar sigma field  involved in the model. The strong electric field region is a subtle case since Schwinger instabilities associated to pair productions of pions could appear. This scenario goes beyond the present discussion and will be considered in a future analysis.

 Our discussion also consider for the first time  box diagrams, generated when sigma propagators are included. In previous analysis those diagrams were not taken into account   \cite{scattering1,scattering2,scattering3} due the the relative high mass of the sigma field.  Although the influence of those diagrams, as expected, is in fact small, they certainly play a relevant role when summing up ladder diagrams. As we know, from the resumation of ladder diagrams reggeized amplitudes emerge which here might depend on external agents.

This analysis of $\pi$-$\pi$ scattering lengths corrections has been carried out previously, using the same model, for the magnetic case \cite{scattering1,scattering2,scattering3}. We found that electric field corrections turn out to be opposite respect to the equivalent magnetic corrections. This is an indication that it will be not an easy task to isolate and identify the influence of certain specific external agents on physical observables.

\section{Linear sigma model and $\pi$-$\pi$ Scattering lengths}

Gell-Mann and Lévy \cite{Gell-Mann:1960mvl} proposed the linear sigma model (LSM) as a framework to elucidate chiral symmetry breaking through both explicit and spontaneous mechanisms. In the chiral broken phase the model is expressed as
\begin{eqnarray}
&&\mathcal{L}=\bar{\psi}\left[i\gamma^{\mu}\partial_{\mu}-m_{\psi}-g(\sigma+i\vec{\pi}\cdot\vec{\tau}\gamma_{5})\right]\psi \nonumber \\
&&+\frac{1}{2}\left[(\partial\vec{\pi})^2+m_{\pi}^2\vec{\pi}^2\right]+\frac{1}{2}\left[(\partial\sigma)^2+m_{\sigma}^2 \sigma^2\right]\nonumber\\
&&-\lambda^2v\sigma(\sigma^2+\vec{\pi}^2)-\frac{\lambda^2}{4}(\sigma^2+\vec{\pi}^2)^2+ \nonumber\\
&&(\varepsilon c-vm_{\pi}^2)\sigma.
\end{eqnarray}
Pions correspond to an isospin triplet, $\vec{\pi}=(\pi_1,\pi_2,\pi_3)$, $c\sigma$  breaks explicitly the $SU(2) \times SU(2)$ chiral symmetry, being $\sigma$ a scalar field, and $\varepsilon$  a small dimensionless parameter. The model also incorporates a doublet of Fermi fields, associated in the original version of the model to nucleon states, which in our context  will be ignored since they are too heavy as compared to the scalar sigma meson mass and to the relevant energy scale. It is interestin  to note that the masses of all fields in the model are determined by $v$. Indeed, the following relations can be proved to be valid:
$m_{\psi}=gv$, $m_{\pi}^2=\mu^2+\lambda^2v^2$ and
$m_{\sigma}^2=\mu^2+3\lambda^2v^2$. 
Perturbation theory at the tree level allows us to identify the pion decay constants as $f_{\pi}=v$. 

The LSM has been a powerful scenario for exploring the effects of external agents like temperature, magnetic field, electric field, and vorticity. These consequences have been studied in a series of articles by various authors, ~\cite{wagner,kovacs2,kovacs1,kovacs3}. In the present work we will explore, in the frame of the LSM model, how an external electric field, generated in collisions between  heavy and  light nuclei, as for example  Au-Cu collisions, will affect the $\pi$-$\pi$ scattering lengths. We will compare our results with previous analysis where a magnetic field was considered.

The most general decomposition for the scattering amplitude for particles with definite isospin quantum numbers is given by~\cite{Collins}

\begin{eqnarray}
T_{\alpha\beta;\delta\gamma}&=& A(s,t,u)\delta_{\alpha\beta}\delta_{\gamma\epsilon}+A(t,s,u)\delta_{\alpha\epsilon}\delta_{\beta\gamma}\nonumber\\
&&+A(u,t,s)\delta_{\alpha\gamma}\delta_{\beta\epsilon},
\label{proyectores}
\end{eqnarray}

\noindent where $\alpha$, $\beta$, $\gamma$, $\delta$ represent isospin components.

 Through suitable projection operators
\begin{align}
P_0&=\frac{1}{3}\delta_{\alpha\beta}\delta_{\gamma\epsilon}\label{ProjOp0},\\
P_1&=-\frac{1}{2}(\delta_{\alpha\gamma}\delta_{\beta\epsilon}-\delta_{\alpha\epsilon}\delta_{\beta\gamma})\label{ProjOp1},\\
P_2&=\frac{1}{2}(\delta_{\alpha\gamma}\delta_{\beta\epsilon}+\delta_{\alpha\epsilon}\delta_{\beta\gamma}-\frac{2}{3}\delta_{\alpha\beta}\delta_{\gamma\epsilon})\label{ProjOp2},
\end{align}
it is possible to find the following isospin dependent scattering amplitudes

\begin{align}
T^{0}&=3A(s,t,u)+A(t,s,u)+A(u,t,s),\label{eq3}\\
T^{1}&=A(t,s,u)-A(u,t,s),\label{eq4}\\
T^{2}&=A(t,s,u)+A(u,t,s),
\label{eq5}
\end{align}

\noindent where $T^I$ denotes a scattering amplitude in a given isospin channel $I = \{0,1,2\}$.\\

As it is well known ~\cite{Collins}, below the inelastic threshold any scattering amplitude can be expanded in terms of partial amplitudes parameterized by phase shifts for each angular momentum channel $\ell$. Hence, in the low-energy region the isospin dependent scattering amplitude can be expanded in partial wave components $T_\ell^I$. The real part of such an amplitude
\begin{equation}
\Re\left(T_{\ell}^{I}\right)=\left(\frac{p^{2}}{m_{\pi}^{2}}\right)^{\ell}\left(a_{\ell}^{I}+\frac{p^2}{m_{\pi}^{2}}b_{\ell}^{I}+\ldots\right),
\end{equation}
is normally expressed in terms of the
scattering lengths $a_{\ell}^{I}$, and the scattering slopes $b_{\ell}^{I}$, respectively.
The scattering lengths satisfy the hierarchy $|a_{0}^{I}|>|a_{1}^{I}|>|a_{2}^{I}|...$.
Specifically, in order to obtain the scattering lengths $a_0^I$, it is
sufficient to calculate the scattering amplitude $T^I$ in the static
limit, i.e., when $s \to 4m_\pi^2$, $t\to 0$ and $u\to 0$,
\begin{equation}
a_{0}^{I}=\frac{1}{32\pi}T^{I}\left(s \to 4m_{\pi}^2,t\to 0, u\to0\right).\label{eq:a0I}
\end{equation} 
 The first measurement of $\pi $-$\pi $ scattering lengths was carried on by Rosellet et al.
~\cite{Rosselet}. More recently, these parameters have been measured  using pionium atoms in the DIRAC experiment \cite{adeva}, as well as through the decay of heavy quarkonium states into $\pi $-$\pi $ final states, where the so called cusp-effect was found~\cite{Liu:2012dv}. 

\section{\label{sec3}Scattering lengths at finite electric field}

In previous works we have calculated the thermo-magnetic dependence of the $\pi$-$\pi$ scattering lengths within the framework of the linear sigma model \cite{scattering1,scattering2,scattering3}. Here we will employ the same model, focusing on the  electric field  dependence of the $\pi$-$\pi$ scattering lengths. For this purpose, we will use the charged bosonic scalar propagator in the presence of an electric field \cite{propagadorelectrico1,propagadorelectrico2}, given by

\begin{equation}D(p) =  \int _{0} ^{\infty}{ds}\frac{e^ {-s\left(\frac{\tanh (qiEs)}{qiEs} p_{\parallel}^2 +p_{\perp}^2+ m^2 \right)}}{\cosh (qiEs)}, \label{propE}
\end{equation} 

\noindent
where $q$ is the electric charge, $p_{\parallel}$ and $p_{\perp}$
refer to $(p_{4},0, 0, p_{3})$ and $(0, p_{1},p_{2},0)$, respectively. For simplicity, the electric field points along the z-axis. Note that in the euclidean version $p^{2} = p_{\parallel}^2 + p_{\perp}^2$=$p_4^2+p_3^2+p_1^2+p_2^2$. 
We are interested in the weak electric field region, to avoid the possible generation of  charged pion pairs $\pi_{\pm}$ that might appear in the strong electric field region through the Schwinger effect.

 Therefore we will proceed to expand the previous expression up to order $\mathcal{O}(E^2)$, to obtain

 \begin{eqnarray}
&&D(p)\approx \frac{1}{p^2+m^2} \nonumber \\
&&-(qE)^2\left(-\frac{1}{(p^2+m^2)^3}+\frac{2  p_{\parallel}^2}{(p^2+m^2)^4}\right).
\label{PiPropagator}
\end{eqnarray}
However, using the relation $p^2=p_\parallel^2+p_\perp^2$, Eq.~\eqref{PiPropagator} can be written as 
\begin{align}
D(p)&\approx\dfrac{1}{p^2+m^2}+\dfrac{q^2E^2\qty[2\qty(p_\perp^2+m^2)-\qty(p^2+m^2)]}{\qty(p^2+m^2)^4}.
\label{PiPropagatorModificado}
\end{align}
The above expression, being more symmetric, is useful to carry on  the integrals that will appear when computing all relevant loop corrections. Also, to distinguish between the free and charged propagators, we will define the free scalar propagator as
\begin{equation}
S(p)=\dfrac{1}{p^2+m^2}.
\label{libre}
\end{equation}

\subsection{\label{sec3a}Loop Integrals Classification}
\begin{figure}[hbtp]
\centering
\subfloat[]{\includegraphics[scale=0.35]{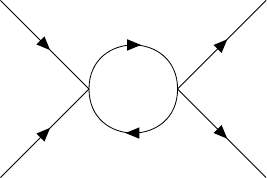}} \hspace*{6mm}
\subfloat[]{\includegraphics[scale=0.35]{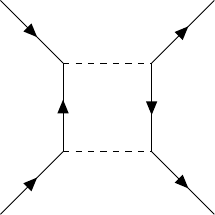}} \hspace*{6mm}
\subfloat[]{\includegraphics[scale=0.35]{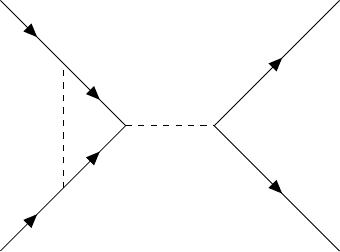}} \\
\subfloat[]{\includegraphics[scale=0.35]{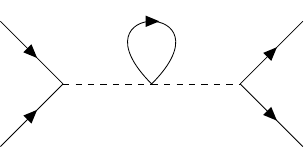}} \hspace*{6mm}
\subfloat[]{\includegraphics[scale=0.35]{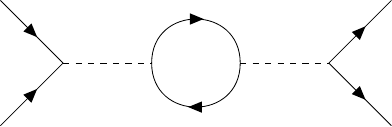}} \hspace*{6mm}
\subfloat[]{\includegraphics[scale=0.35]{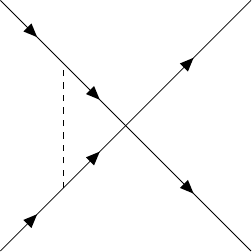}} \\
\subfloat[]{\includegraphics[scale=0.35]{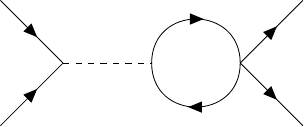}}\hspace*{6mm}
\subfloat[]{\includegraphics[scale=0.35]{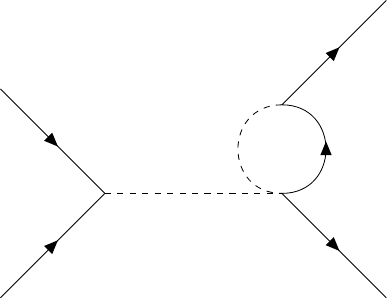}} \hspace*{6mm}
\subfloat[]{\includegraphics[scale=0.35]{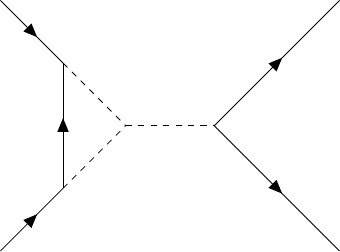}} \\
\subfloat[]{\includegraphics[scale=0.35]{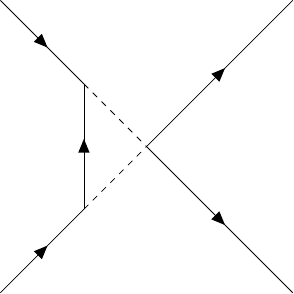}}
\caption{$s$ channel diagrams.}
\label{fig:schanneldiagrams}
\end{figure}

\begin{figure}[hbtp]
\centering
\subfloat[]{\includegraphics[scale=0.35]{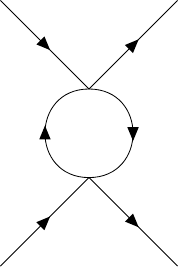}} \hspace*{8mm}
\subfloat[]{\includegraphics[scale=0.35]{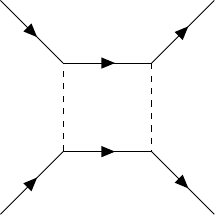}} \hspace*{8mm}
\subfloat[]{\includegraphics[scale=0.35]{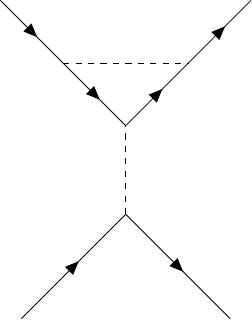}} \hspace*{8mm}
\subfloat[]{\includegraphics[scale=0.35]{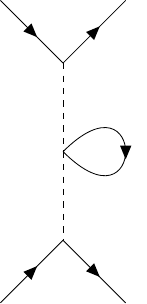}} \\
\subfloat[\textcolor{red}{}]{\includegraphics[scale=0.35]{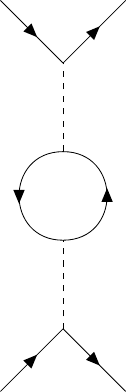}} \hspace*{8mm}
\subfloat[]{\includegraphics[scale=0.35]{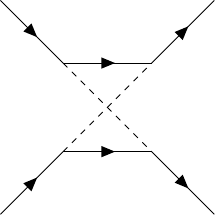}} \hspace*{8mm}
\subfloat[]{\includegraphics[scale=0.35]{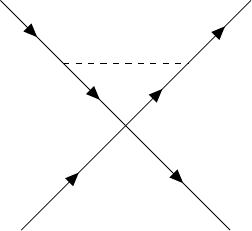}}\hspace*{8mm}
\subfloat[]{\includegraphics[scale=0.35]{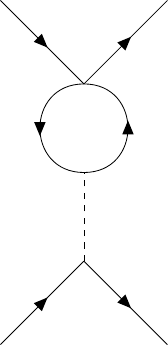}} \\
\subfloat[]{\includegraphics[scale=0.35]{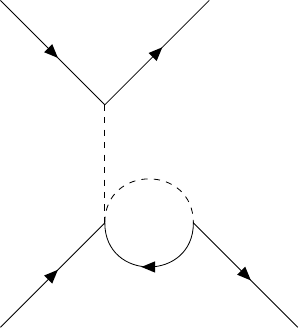}} \hspace*{8mm}
\subfloat[]{\includegraphics[scale=0.35]{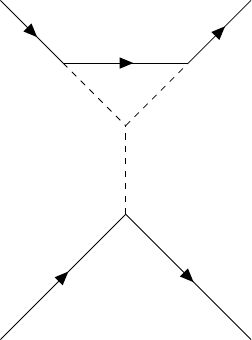}}\hspace*{8mm}
\subfloat[]{\includegraphics[scale=0.35]{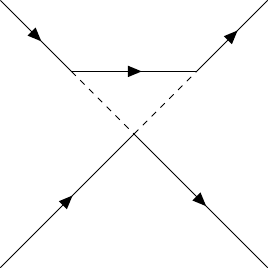}}
\caption{$t$ channel diagrams.}
\label{fig:tchanneldiagrams}
\end{figure}

For the analysis, we need to compute 21 Feynman diagrams. The diagrams that contribute to the $s$ channel are shown in figure \ref{fig:schanneldiagrams}, and those diagrams relevant for  the $t$ channel can be seen in figure \ref{fig:tchanneldiagrams}. The continuous line/ dashed line represents charged pions/$\sigma$-meson. The $u$ channel diagrams are analogous to the $t$ channel. The only difference corresponds to a permutation of isospin indexes of the external legs.

In previous works \cite{scattering1,scattering2,scattering3}, and because the sigma boson mass is much bigger than the pion mass, considering also the static limit approximation, the diagrams that contained sigma bosons were pinched, i.e., the sigma propagator was contracted to a point. The following limit was used
\begin{equation}
\dfrac{1}{p^2+m_\sigma^2} \longrightarrow \dfrac{1}{m_\sigma^2}.
\end{equation}
Pinching the sigma propagators simplify several diagrams, and all the calculations can be reduced to computing only five integrals. 

Due to lack of space, here we will not present any detail concerning the evaluation of the different diagrams. The reader can consult the technical details in the original article\cite{Nosotros24}. As we said previously, one of the novelties is the calculation of the different box diagrams which have not been considered previously. It is found that the scatering amplitude get the form

\subsection{\label{sec3c}Isospin Projections}
Because of the associated Feynman rules, all the integrals emerging from the diagrams have a determined isospin structure. These structures are simplified when isospin projection operators acting on the different integrals are used. Using the projection operators \eqref{ProjOp0}, \eqref{ProjOp1}, and \eqref{ProjOp2},
all projections can be easily obtained. In Fig. 3 we show the evolution of the scattering lengths in the isospin channels $I=0$ and $I=2$. There are no contribution in the $I =1$ channel as it is explained in \cite{Nosotros24}.

\begin{figure}[H]
\centering
\includegraphics[scale=0.5]{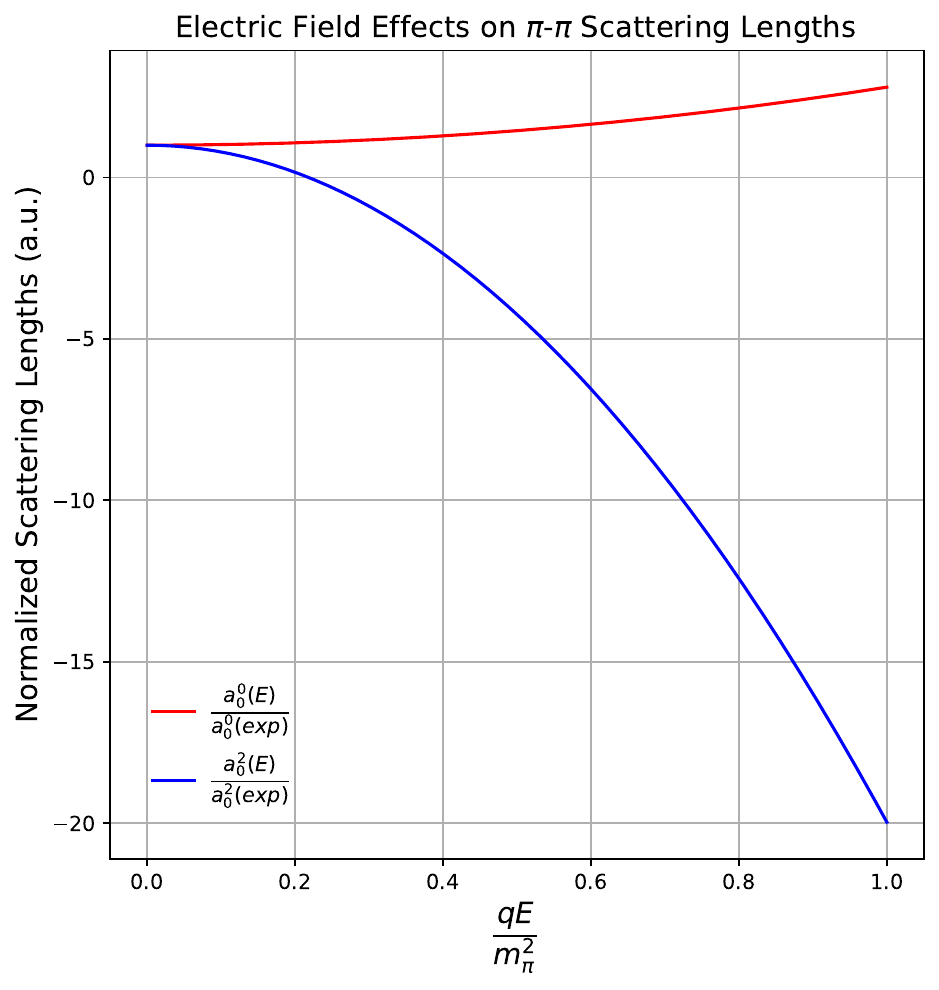}
\caption{Behaviour of the normalized scattering lengths as a function of $qE/m_\pi^2$. Red lines represent $a_0^0(E)/a_0^0(\text{exp})$, while blue lines represent $a_0^2(E)/a_0^2(\text{exp})$.}
\label{fig:Plot}
\end{figure}

In order to discuss the behavior of scattering lengths in the presence of an electric field, we normalize our results by the experimental values at $E = 0$. We will use the following parameters $m_{\pi}=\SI{140}{\mega\electronvolt}$, $m_\sigma=\SI{550}{\mega\electronvolt}$, $v=\SI{89}{\mega\electronvolt}$, and $\lambda^2=4.26$, obtaining the following plot \ref{fig:Plot}.

\section{Conclusions}
We have presented a complete calculation at the one loop level of the corrections induced by an external electric field on the $\pi$-$\pi$ scatering lengths in the frame of the linear sigma model. Our main result is presented in Fig.3. From this plot, we see that $a_0^0(E)$ increases with $a_0^2(E)$ decreases. This behavior is exactly opposite to what occurs in the case of an external magnetic field. Such opposition between electric and magnetic fields is also observed in the discussion of renormalons in the $\lambda \phi^4$ theory \cite{propagadorelectrico1}. Another interesting comparison with the effects of the magnetic field is that the modification of the amplitudes $a_0^2$ with an electric field is much more intense than with a magnetic field. This is related to the different structure of both propagators.  A significant novelty of this calculation is our inclusion of box diagrams, which have been fully computed, unlike in previous works \cite{scattering1,scattering2,scattering3}.Nevertheless, it is worth mentioning that the approximation made in Ref. \cite{scattering1,scattering2,scattering3} is very accurate when compared to the full calculation of the box diagrams. Although the contribution from the box diagrams is small, these diagrams are interesting from a different point of view. In fact, we can buil ladder diagrams whose building blocks are the box diagrams. The re-summation of ladder diagrams might produce a Regge type behavior for the scattering amplitudes. Finally, it would be interesting to explore how temperature effects, together with the presence of an electric field, might change or the results presented here. We plan to carry on this analysis in the near future

\section{Acknowledgements}
M.L., and R.Z. acknowledge support from ANID/CONICYT FONDECYT Regular (Chile) under Grants No. 1220035 and 1241436.

\bibliographystyle{elsarticle-num}
\bibliography{ourbibliography}

\end{document}